\documentclass[10pt,twocolumn,twoside]{IEEEtran}
\IEEEoverridecommandlockouts
\usepackage{cite}
\usepackage{algorithmic}
\usepackage{textcomp}
\usepackage{xcolor}

\usepackage{graphicx,color}
\usepackage{amsfonts,amsthm,amssymb}
\usepackage{mathrsfs,amsmath,arydshln,latexsym}
\usepackage{cite,url}
\usepackage[bookmarks,colorlinks]{hyperref}
\usepackage[linesnumbered,ruled,vlined]{algorithm2e}
\usepackage{multirow,array,makecell}
\usepackage{stfloats}
\usepackage{balance}
\usepackage{fancyhdr}
\usepackage{bm}
\usepackage{enumitem}
\usepackage{pifont}
\usepackage{subcaption}
\usepackage{nomencl}
\usepackage{colortbl}
\usepackage{diagbox}
\usepackage{balance}
\usepackage{threeparttablex}

\allowdisplaybreaks[4]

\newcommand{\tabincell}[2]{\begin{tabular}{@{}#1@{}}#2\end{tabular}}

\begin{document}

	\title{Near-Field Channel Modeling for\\ Holographic MIMO Communications
	}

	\author{Tierui Gong,~\IEEEmembership{Member,~IEEE,} 
		Li Wei,~\IEEEmembership{Member,~IEEE,}
		Chongwen Huang,~\IEEEmembership{Member,~IEEE,}\\ 
		George C. Alexandropoulos,~\IEEEmembership{Senior Member,~IEEE,}
		Mérouane Debbah,~\IEEEmembership{Fellow,~IEEE,}\\  
		and Chau Yuen,~\IEEEmembership{Fellow,~IEEE} 
		\vspace{-0.8cm}
        \thanks{T. Gong, L. Wei and C. Yuen are with the School of Electrical and Electronics Engineering, Nanyang Technological University, Singapore 639798 (e-mail: trgTerry1113@gmail.com, l\_wei@ntu.edu.sg, chau.yuen@ntu.edu.sg).}
		\thanks{C. Huang is with the College of Information Science and Electronic Engineering, Zhejiang University, Hangzhou 310027, China (e-mails: chongwenhuang@zju.edu.cn).}
		\thanks{G. C. Alexandropoulos is with the Department of Informatics and Telecommunications, National and Kapodistrian University of Athens, 15784 Athens, Greece (e-mail: alexandg@di.uoa.gr).}
		\thanks{M. Debbah is with Department of Electrical Engineering and Computer Science, Khalifa University of Science and Technology, Abu Dhabi 127788, United Arab Emirates, and also with the CentraleSupelec, University ParisSaclay, 91192 Gif-sur-Yvette, France (e-mail: merouane.debbah@ku.ac.ae).}
	}

	\maketitle

\begin{abstract} 
Empowered by the latest progress on innovative metamaterials/metasurfaces and advanced antenna technologies, holographic multiple-input multiple-output (H-MIMO) emerges as a promising technology to fulfill the extreme goals of the sixth-generation (6G) wireless networks. The antenna arrays utilized in H-MIMO comprise massive (possibly to extreme extent) numbers of antenna elements, densely spaced less than half-a-wavelength and integrated into a compact space, realizing an almost continuous aperture. Thanks to the expected low cost, size, weight, and power consumption, such apertures are expected to be largely fabricated for near-field communications. In addition, the physical features of H-MIMO enable manipulations directly on the electromagnetic (EM) wave domain and spatial multiplexing. To fully leverage this potential, near-field H-MIMO channel modeling, especially from the EM perspective, is of paramount significance. In this article, we overview near-field H-MIMO channel models elaborating on the various modeling categories and respective features, as well as their challenges and evaluation criteria. We also present EM-domain channel models that address the inherit computational and measurement complexities. Finally, the article is concluded with a set of future research directions on the topic. 
\end{abstract}

\section{Introduction}
The future sixth-generation (6G) of wireless networks is expected to support a wide variety of scenarios (space, air, ground, and sea communications), functionalities (communication, positioning, sensing, computing, and imaging), and emerging applications (e.g., multisensory extended reality and holographic videos), which require advanced capabilities to foster high spectral and energy efficiencies, extremely massive connectivity, and ultra-low end-to-end latency. Unfortunately, today's wireless technologies, e.g., massive multiple-input multiple-output (M-MIMO), fail to satisfy such extreme requirements in a sustainable way. On the other hand, recent advances on metamaterials and metasurfaces as well as on antenna technologies gave birth to the holographic MIMO (H-MIMO) concept that emerges recently as a 6G promising technology with significant potential to enable a multitude of new functionalities and upper layer applications \cite{Gong2023Holographic}. 
	
The antenna aperture utilized in H-MIMO comprises a plurality of densely distributed antenna elements, whose inter-element spacing is less than half of the wavelength of the impinging electromagnetic (EM) wave. Those elements are usually implemented via metamaterials, metasurfaces, or tightly coupled antennas, which are capable of controlling EM waves with various expected responses, and, as such, the resulting antenna aperture appears to be nearly continuous. This distinctive feature of H-MIMO offers powerful wave control capability, allowing manipulation of EM waves in unprecedented levels (e.g., recording and reconstruction of the completed vector wave field is feasible in a nearly continuous space), thus, paving the way to sophisticated signal processing techniques realized entirely in the EM domain \cite{Gong2023Holographic}. 
	
In addition, the antenna apertures realized via advanced metamaterials and metasurfaces exhibit low power consumption and can be fabricated with low cost, which implies that they can be designed to be electrically large, having extremely massive numbers of antenna elements packed over them. Such electrically large apertures facilitate wireless operations in the near field, implying that the communication distance between a transmitter (TX) and a receiver (RX) falls within the Fresnel region \cite{Zhang2022Beam}, in contract to conventional operations which take place in the far field. Within this region, EM wave propagation exhibits spherical wavefronts instead of planar ones. Interestingly, near-field communications provide additional spatial degrees-of-freedom (DoF) over far-field communications even in line-of-sight (LoS) channel conditions, which facilitates extremely large spatial multiplexing. 
	
To leverage the potential of H-MIMO which will mainly operate in the near field, the fundamental aspects of near-field EM wave propagation need to be well understood. To this end, accurate near-field channel modeling will facilitate unveiling the fundamental limits of wireless operations in this region, but will also enable efficient H-MIMO system designs. The dense and large-size characteristics of H-MIMO bring new challenges in near-field channel modeling that need to be adequately addressed. In this article, we aim to provide a panoramic reference to the near-field H-MIMO channel modeling for both industry and academia, introducing efficient EM-domain channel models for the emerging H-MIMO paradigm. To the best of our knowledge, this is the first article contributing such a comprehensive overview to the area. Apart from a straightforward survey of existing works, we discuss the latest highly-organized near-field channel modeling categories and present an in-depth generalization of their distinctive features, modeling challenges, and evaluation criteria, as well as a list of key research directions on the topic.
	
\section{H-MIMO Channel Modeling}
\label{Section_CCM}
Recent advances on metasurface-based antenna apertures enable H-MIMO. Such apertures can be active or passive, transmissive or reflective, and can be fed by waveguide structures or external sources~\cite{Gong2023Holographic}. In the following, we present the most significant categories of H-MIMO channel modeling.

\subsection{Communication Distance}
The communication distance between transceivers determines if their wireless link is realized in the near- of far-field. In fact, the propagation characteristics of EM waves are totally different in these two regions. Normally,the criterion for empirically discriminating the near- from the far-field region is the Rayleigh distance \cite{Zhang2022Beam}. In the far field, it is indicated that the communication distance exceeds the Rayleigh distance, while conversely, the communication distance falls within the Rayleigh distance in the near field (see Fig.~\ref{fig:ModelingCategory}(a)).

\subsubsection{Far-Field Modeling (FFM)}
The far field is also known as the Fraunhofer region, where the propagation wavefront of traveling EM waves is a plane. In this region, the path loss is inversely proportional to the square of the link distance. It is remarkable that the far field is mostly considered in conventional M-MIMO systems. In this regard, the EM waves seen by different M-MIMO antennas are approximately equal in signal amplitudes, while differing in signal phases. Due to the plane wavefront, the phase differences can be feasibly interpreted by the indices of the antennas and their relative spacing. As such, the far-field channel model for antenna array systems can be depicted via the array response vector, which depends on the waves' angle of arrival/departure (AoA/AoD). For example, the multipath channel realized via the cluster-based approach in \cite{Elbir2023Twenty} is built upon the array response vector.
	
\subsubsection{Near-Field Modeling (NFM)}
For a given communication link distance, the larger are the deployed antenna apertures, the more EM wave propagation happens closer to the near field. In this region, the EM propagation wavefront becomes spherical. In fact, the near field can be divided into the reactive and the radiative near field (also known as the Fresnel region (see Fig.~\ref{fig:ModelingCategory}(a)). The propagation energy can deflect back to the source in the former region, while it fails in the latter region. The path losses in the reactive and radiative near-field regions are inversely proportional to the sixth and fourth power of the communication distance, respectively. In addition, in view of the spherical wavefront in the near field, the EM waves captured by different antennas experience different signal amplitudes and phases due to different distances from the signal source. This implies that the angle-dependent array response vector in the far field becomes distance-angle-dependent in the near field. The free-space near-field channel models in \cite{Zhang2022Beam,Do2023Parabolic} describe mathematically the spherical and parabolic wavefronts.

\subsection{Existence of Scatterers}
Channel models can be also taxonomized into line-of-sight (LoS) and non-line-of-sight (NLoS) ones, depending on if scatterers exist in the propagation environment, as shown in Fig.~\ref{fig:ModelingCategory}(b).

\subsubsection{Line-of-Sight (LoS)}
The LoS channel model describes free-space transmissions without any scatterers. It will become more and more important in future near-field wireless systems equipped with extremely large antenna apertures, and due to the shifting of operating frequencies from low frequency bands to THz. In a LoS scenario, wave propagation experiences different path losses in different communication distances, e.g., far or near field. In antenna array systems, the LoS channel model is expressed via the angle-dependent or distance-angle-dependent array response vectors depending on the communication distance \cite{Zhang2022Beam}. In continuous antenna aperture systems, the LoS channel model is established via the tensor Green's function (TGF) \cite{Dardari2020Communicating2,Gong2023HMIMO,Gong2023Transmit} that provides physical consistency.

\subsubsection{Non-Line-of-Sight (NLoS)}
On the contrary, the NLoS channel model includes multiple indirect propagation paths generated when waves pass through a certain number of scatterers. As a widely used NLoS channel model, the multipath channel, usually created via the cluster-based approach, considers each path between the TX (or RX) array and each scatterer as a free-space LoS path, which, in antenna array systems, exhibits an angle-dependent (far field) or distance-angle-dependent (near field) array response vector. Therefore, the NLoS channel is represented as a sum of the array response vectors over the scatterers and the transceiver antenna arrays~\cite{Elbir2023Twenty,Cui2022Channel}. Alternatively~\cite{Dong2022Near}, NLoS conditions can be modeled via the second-order channel statistics, i.e., the spatial correlation matrix. According to this modeling approach, the product of the array response vector and its conjugate transpose, under some normalized spatial scattering function, is integrated over the propagation azimuth/elevation angles.

\subsection{Parameters' Randomness}
The available channel models can be categorized into deterministic and stochastic as follows.

\subsubsection{Deterministic Modeling (DM)}
Deterministic channel models are generally based on the EM wave theory, according to which wave propagation follows Maxwell's equations. In general, such models are capable of offering highly accurate results. However, they require the exact information of the propagation environment and the locations of the transceivers. In addition, such deterministic models are site-specific, and thus, computationally demanding due to extensive integrals and/or differentials computations appearing in Maxwell's equations that need to be solved. On the other hand, the impulse responses captured by the deterministic models facilitate system design of a detailed level, e.g., network planning and system management. A typical deterministic model is ray tracing \cite{Han2015Multi-Ray} (see Fig.~\ref{fig:ModelingCategory}(c)).

\subsubsection{Stochastic Modeling (SM)}
\begin{figure*}[t!]
		\centering
		\includegraphics[height=11cm, width=16.6cm]{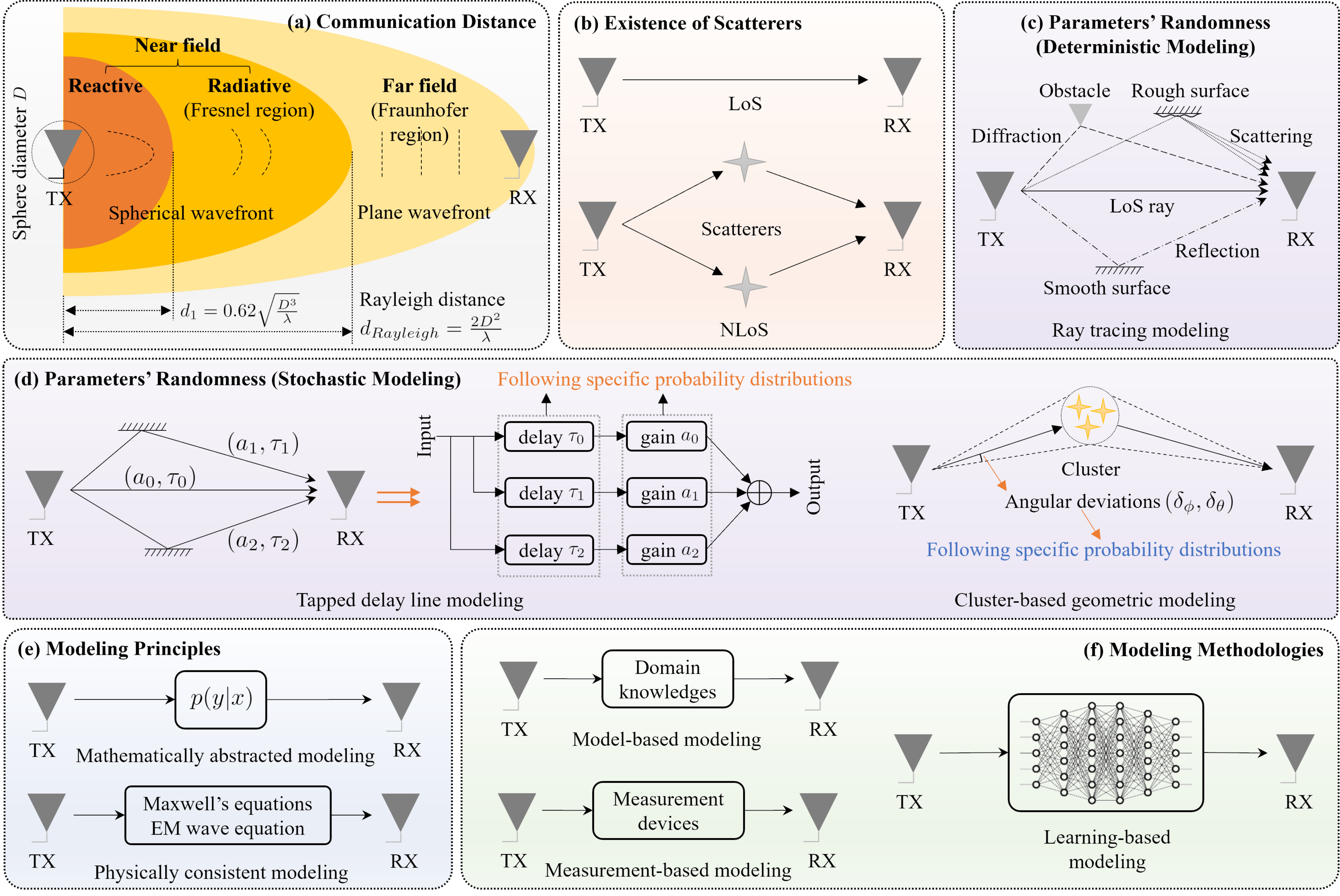}
		\caption{Categories of H-MIMO channel modeling.}
		\label{fig:ModelingCategory}
		\vspace{-1em}
	\end{figure*}

Those channel models exploit certain probability distributions of the channel parameters, providing an effective trade-off between accuracy, computational complexity, flexibility, and mathematical tractability, as compared with the deterministic models. Stochastic models are mostly applied to system designs of a conceptual level. For instance, the multipath channel can be depicted via the tapped delay line with each tap generated following a specific probability distribution \cite{3GPP2017Study}. In addition, the angular spread of AoAs/AoDs \cite{Elbir2023Twenty,Cui2022Channel,3GPP2017Study}, characterizing the multipath propagation direction, is usually assumed to follow a specific probability distribution, e.g., uniform, Gaussian, Laplace, or von Mises distributions. As a consequence, different scattering environments can be mathematically described. Both latter cases are demonstrated in Fig.~\ref{fig:ModelingCategory}(d).

	\subsection{Modeling Principles}
Complying with different principles, channel modeling can be microscopically categorized as mathematically abstracted and physically consistent, as depicted in Fig.~\ref{fig:ModelingCategory}(e).

\subsubsection{Mathematically Abstracted Modeling (MAM)}
MAM implies that the physical channel is modeled to a certain degree in a mathematically abstract relationship between the transmit and receive signals. For example, the channel in Shannon's information theory is treated as a conditional probability distribution. Moreover, the Rayleigh fading channel model is a mathematically abstracted and well fitted model for rich scattering environments. Similarly, the multipath channel model views the wave propagation from TX to RX with multiple propagation paths, where each path is mathematically modeled under a specific probability distribution \cite{Elbir2023Twenty,Cui2022Channel,3GPP2017Study}. These channel models appear to be mathematically tractable and are convenient for system analyses and designs. However, this category ignores certain physical effects of wave propagation, and hence, may fail to fully capture the essence of physical channels.

\subsubsection{Physically Consistent Modeling (PCM)}
To consider the physical effects of wave propagation, i.e., the EM wave propagation characteristics, one can build a physically consistent channel model using fundamental EM properties. In this way, the interactions between the EM wave field and propagation media can be well characterized. Furthermore, fundamental limits, e.g., DoF and channel capacity, can be more accurately revealed compared to MAM models.	As previously mentioned, the highly accurate deterministic channel models, that rely on the EM wave theory, are physically consistent. The recently introduced Fourier plane-wave series expansion channel model in \cite{Pizzo2022Fourier} and the TGF-based channel model in \cite{Dardari2020Communicating2} are physically consistent for far-field NLoS and near-field LoS channels, respectively, allowing precise analysis of wireless systems in the EM domain.

\subsection{Modeling Methodologies}
Channel modeling methods can be divided into model-based, measurement0based, and learning-based, as shown in Fig. \ref{fig:ModelingCategory}(f) and discussed below.

\subsubsection{Model-Based Modeling (MoBM)}
The channel models belonging to this category are analytical with well established mathematical formulas, which are capable of describing a variety of wireless channels, more commonly, the channel impulse responses and/or its stochastic characteristics. The model-based category offers an effective and convenient way for system analyses and designs, allowing fast performance evaluations on a conceptual level.

\subsubsection{Measurement-Based Modeling (MeBM)}
This category relies on sophisticated measurement devices, such as a channel sounder \cite{Gentile2018Millimeter}. With such devices, a large variety of channel parameters, e.g., AoAs/AoDs, time delays, Doppler shifts, and complex path gains, can be obtained. The measurement-based models provide a relatively high accuracy, but are, however, device-specific and -dependent, time-consuming, and limited to the specific measurement environment, or at best, to similar ones. Interestingly, measurements can be used to verify the prediction quality of deterministic channel models. Moreover, the parameters of stochastic channel models can be obtained through channel measurements.

\subsubsection{Learning-Based Modeling (LBM)}
LBM is a recent trend, with a representative work deploying a generative adversarial network  to model the wireless channel~\cite{Yang2019Generative}. This category allows to fit more complicated data distributions to the propagation of the complex wave field, thus, facilitating more accurate fitting. Furthermore, LBM is capable of relieving the requirements of specific domain knowledge, which is needed in conventional non-learning-based channel models.

\section{Near-Field H-MIMO Channel Models}
\label{Section_NFCM}
In this section, we present channel modeling aspects for near-field H-MIMO systems. We commence by describing the main features and challenges with near-field H-MIMO channel modeling, and then, discuss evaluation criteria for such channel models. We also overview several existing near-field H-MIMO channel models. 
	
\subsection{Distinctive Features}
EM wave propagation in near-field H-MIMO systems comes with various new features, which have not being well treated in the widely studied far-field M-MIMO channel modeling. Those new features are mainly attributed to the `large' (features \textbf{F1}, \textbf{F3}, \textbf{F5}, \textbf{F6}, and \textbf{F8} that follow), `dense' (features \textbf{F6} and \textbf{F7}), and strong EM wave manipulation (features \textbf{F2} and \textbf{F4}) characteristics of H-MIMO. The recently presented near-field channel models are expected to capture those new features, which are discussed in the sequel.
	
\begin{figure*}[t!]
		\centering
		\includegraphics[height=7.6cm, width=16.6cm]{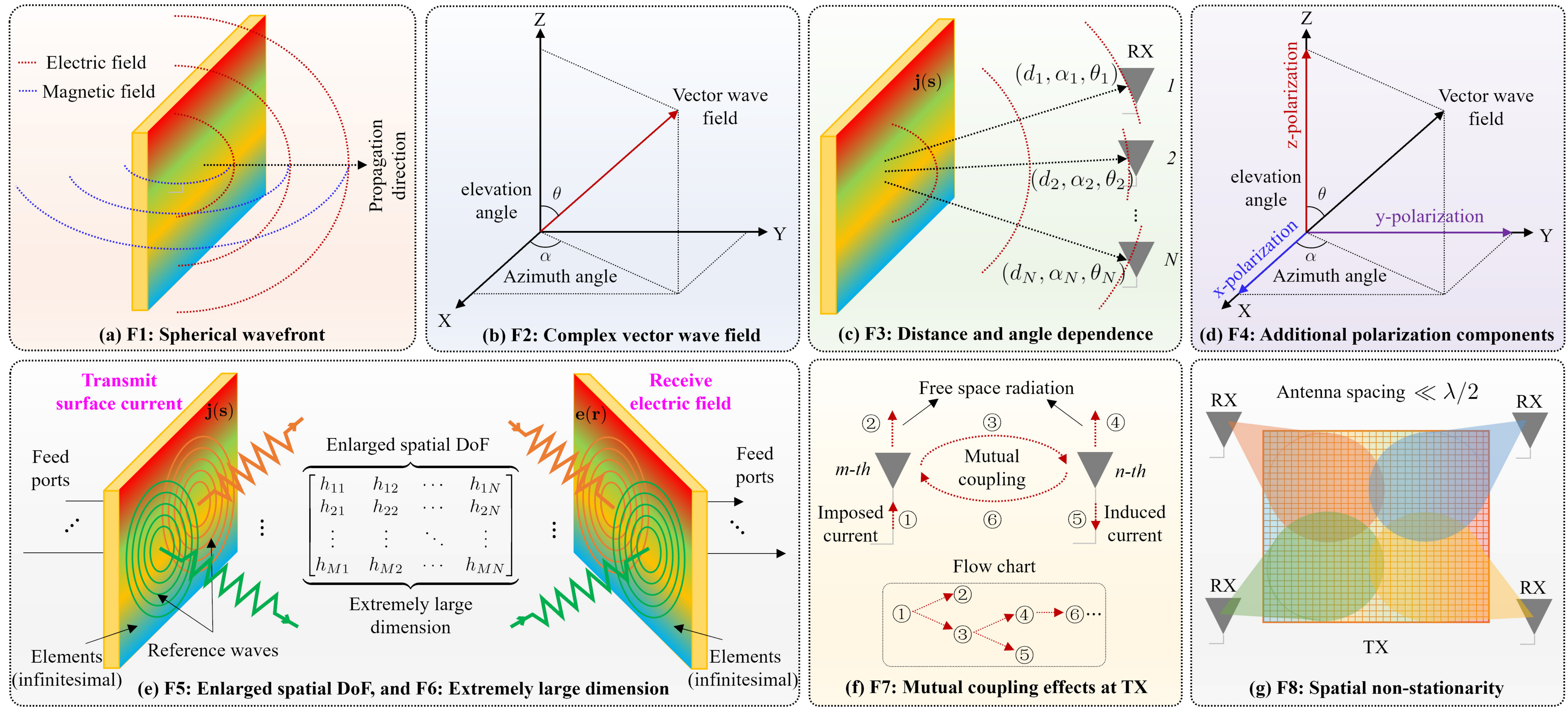}
		\caption{Distinctive features of near-field H-MIMO channels.}
		\label{fig:ChannelFeatures}
		\vspace{-1em}
	\end{figure*}

	\begin{itemize}
		\item 
		\textbf{F1: Spherical wavefront.} 
		Dissimilar to the planar wavefront propagation in far-field channels, EM waves in the near field spread out from the source to outgoing directions in the form of spheres (Fig. \ref{fig:ChannelFeatures}(a)). This implies that the wavefront captured from RX antennas becomes spherical. 
		
		\item 
		\textbf{F2: Complex vector wave field.} 
		The vast majority of far-field channel models considers one-dimensional scalar wave propagation. There exist only few far-field models that elaborate on two-dimensional dual-polarization planar wavefront propagation. Those channel models actually constitute a simple compound of channels for scalar wave field, which fails to accurately describe complex near-field H-MIMO channels. In the near field, EM wave fields are three-dimensional spatial vectors with three orientations, as shown in Fig.~\ref{fig:ChannelFeatures}(b), whose behavior needs to be precisely modeled in near-field H-MIMO systems.
		
		\item 
		\textbf{F3: Distance and angle dependence.}
Due to the spherical wavefront, the RX antennas of a planar aperture, when located in the near field, capture EM waves with distinct amplitudes and phases, depending on both the distance and the angle between the TX and each RX, as illustrated in Fig.~\ref{fig:ChannelFeatures}(c). 
		
		\item 
		\textbf{F4: Additional polarization components.}
		The wave fields in the near field are spatially vectorial and exhibit complex EM behavior, hence, the near-field H-MIMO channel can realize abundant polarization forms (e.g., linear (see Fig. \ref{fig:ChannelFeatures}(d)), circular, and elliptical) with tri-polarization components. This feature is distinctive as compared with dual-polarization channel models for planar wavefront propagation.
		
		\item 
		\textbf{F5: Enlarged spatial DoF.}
		Compared with conventional far-field channels, near-field H-MIMO channels provide additional spatial DoF, as illustrated in Fig.~\ref{fig:ChannelFeatures}(e). This feature is attributed to the enriched distance information induced due to the spherical wavefront, as well as to the additional polarization components.
		
		\item 
		\textbf{F6: Extremely large dimension.}
H-MIMO systems are expected to utilize nearly continuous antenna apertures, implying that the dimensions of the channel matrix will be very large. This situation will become even more severe as the aperture becomes extremely large in the near field (see Fig.~\ref{fig:ChannelFeatures}(e)).
		
		\item 
		\textbf{F7: Mutual coupling effects.}
Due to the densely deployed antenna elements in H-MIMO antenna apertures, namely, the inter-element spacing being far less than half a wavelength, mutual coupling will appear that needs to be accurately modeled and accounted for, as depicted in Fig.~\ref{fig:ChannelFeatures}(f). 

		\item 
		\textbf{F8: Spatial non-stationarity.} 
		For extremely large H-MIMO apertures, spatial non-stationarity will appear in the resulting wireless channels. This feature indicates that some of the RXs will be visible to a portion of the TX antenna aperture, such that different regions of the aperture may experience different propagation environments, as shown in Fig. \ref{fig:ChannelFeatures}(g).
	
	\end{itemize}

	\subsection{Challenges}
It becomes apparent from the previously described distinctive features of near-filed H-MIMO channels that several new design challenge will arise. In the following, we present the most critical of them.

	\begin{itemize}		
		\item 
		\textbf{C1: Vectorial spherical wavefront inclusion.}
		As previously discussed, vector wave field and spherical wavefront are notable features of near-field H-MIMO channels, which need to be adequately captured in the respective channel models. To date, the most spherical wavefront channel models are developed for scalar wave fields. In addition, near-field H-MIMO channel modeling needs to accurately describe polarization effects.
		
		\item 
		\textbf{C2: Concise channel models with manageable dimensions.}
		The dimension of an H-MIMO channel model may reach extremely large values, possibly leading to computational inefficiency and/or costly parameter measurements. To tackle this problem, concise channel models with reduced dimensions are critical, which is quite a challenging task.
		
		\item 
		\textbf{C3: Mutual coupling awareness.}
		In H-MIMO systems, mutual coupling effects cannot be ignored. During transmissions, the radiated wave field of each antenna element influences the radiations of near-by ones, which further affects the radiation subsequently in an infinite loop. Those combined effects influence the propagation channel. The design of mutual-coupling-aware channel models will facilitate the understanding and exploitation of this phenomenon.
		
		\item 
		\textbf{C4: Spatial non-stationarity awareness.}
		In extremely large-scale H-MIMO systems, the spatial non-stationarity needs to be carefully incorporated in the near-field channel models. This feature's precise modeling, which may require a multitude of scenario-specific measurements, needs to drive non-stationarity-aware channel models.

		\item 
		\textbf{C5: Scalability to traditional channel models.}
		In near-field H-MIMO channel modeling, a variety of new features arise. However, those models need to degenerate in far-field M-MIMO channel models that comprise special cases (i.e., for $\lambda/2$-spacing, planar wavefront, etc.).  To this end, scalable channel models incorporating traditional ones are expected, however, this is quite challenging due to the seamless blend of the available different modeling approaches.
	\end{itemize}

\subsection{Evaluation Criteria}
With the various available near-field H-MIMO channel models, a comprehensive comparison among them and between the ground truth becomes challenging. We next list several evaluation criteria for such comparisons.
	
	\begin{itemize}
		\item 
		\textbf{High accuracy (HA):}
High modeling accuracy is of paramount importance for both performance evaluation and system designs. The accuracy refers to the models' precise depiction capability and is generally validated by comparing the designed channel model with realistically measured data and/or theoretical results. Intuitively, for the modeling accuracy comparison of the previously discussed principles and methodologies holds: DM $>$ SM, PCM $>$ MAM, and MeBM $>$ LBM $>$ MoBM.
		
		\item 
		\textbf{Low computational complexity (LCC):}
		For practical implementations, the computational complexity of channel models is an important metric. For H-MIMO systems, this is even more critical, since such models are expected to have extremely large  numbers of parameters. For instance, channel modeling for a continuous H-MIMO aperture need to address the calculation of computationally-demanding integrals. Therefore, the development of low complexity channel models is of great significance.
		
		\item 
		\textbf{Low measurement complexity (LMC):}
		The measurement complexity is another important aspect of channel modeling for practical realizations. A channel model may rely on absolute and/or the relative information (e.g., coordinates and distances/angles) that need to be measured. The extremely large numbers of parameters may cause measurement infeasibility, thereby inspiring the adoption of measurement-efficient channel models.
		
		\item 
		\textbf{High flexibility and mathematical tractability (HFMT):}
		These aspects reveal the generalization and robustness of a channel model to a variety of channel conditions and application scenarios, as well as for facilitating system analyses and designs. Nevertheless, the trade-off between these metrics and accuracy need to be wisely balanced.  
		
	\end{itemize}
	
	\begin{table*}[!t]
		\tiny
		\renewcommand{\arraystretch}{1.2}
		\caption{\textsc{State-of-the-Art Near-field H-MIMO Channel Models.}}
		\label{tab:ChannelModeling}
		\centering
		\tabcolsep = 0.1cm
		\resizebox{\linewidth}{!}{
			\begin{tabular}{!{\vrule width0.6pt}c|c|c|c|c|c|l!{\vrule width0.6pt}}			
				\Xhline{0.6pt}
				\rowcolor{yellow} \textbf{\tabincell{c}{Channel model}} & \textbf{\tabincell{c}{Categories}} & \textbf{\tabincell{c}{Features}} & \textbf{\tabincell{c}{Challenges}} & \textbf{\tabincell{c}{Evaluation}} & \textbf{Ref.} & \qquad \qquad \qquad \qquad \qquad \qquad \qquad \textbf{\tabincell{c}{Main contributions}}\\
				
				\Xhline{0.6pt}
				{\tabincell{c}{Spherical wavefront\\ propagation model}} & {\tabincell{c}{LoS, DM, MAM, MoBM}} & {\tabincell{c}{F1, F3, F5, F6}} & {\tabincell{c}{C1, C3, C4}} & 
				{LCC, LMC, HFMT} & \tabincell{c}{\cite{Zhang2022Beam}} & \tabincell{l}{Near-field channel model for dynamic metasurface antenna systems\\ as well as beam focusing for different antenna structures.} \\
				
				\Xhline{0.6pt}
				{\tabincell{c}{Parabolic wavefront\\ propagation model}} & {\tabincell{c}{LoS, DM, MAM, MoBM}} & {\tabincell{c}{F1, F3, F5, F6}} & {\tabincell{c}{C1, C3, C4}} & 
				{LCC, LMC, HFMT} & \tabincell{c}{\cite{Do2023Parabolic}} & \tabincell{l}{Parabolic wavefront model for LoS channels, study of its validity and applicability,\\ and discussion on its properties and significance to LoS MIMO systems.} \\
				
				\Xhline{0.6pt}
				\multirow{5}{*}{\tabincell{c}{Tensor Green's\\ function model}} & {\tabincell{c}{LoS, DM, PCM, MoBM}} & {\tabincell{c}{F1 --- F6}} & {\tabincell{c}{C2 --- C5}} & 
				{HA} & \tabincell{c}{\cite{Dardari2020Communicating2}} & \tabincell{l}{Employ TGF as the near-field LoS channel model, and study of the fundamental\\ limits of a point-to-point H-MIMO system.} \\
				
				\cline{2-7}
				& {\tabincell{c}{LoS, DM, PCM, MoBM}} & {\tabincell{c}{F1 --- F6}} & {\tabincell{c}{C2 --- C5}} & 
				{HA, LCC} & \tabincell{c}{\cite{Gong2023HMIMO}} & \tabincell{l}{Development of computationally efficient EM-domain LoS channel models based\\ on the TGF and study of the capacity limit of a point-to-point H-MIMO system.} \\
				
				\cline{2-7}
				& {\tabincell{c}{LoS, DM, PCM, MoBM}} & {\tabincell{c}{F1 --- F6}} & {\tabincell{c}{C3, C4}} & 
				\tabincell{c}{HA, LCC, LMC,\\ HFMT} & \tabincell{c}{\cite{Gong2023Transmit}} & \tabincell{l}{Development of measurement efficient EM-domain LoS channel models based\\ on the TGF, which realizes the separability of coupled TX-RX parameters.} \\
				
				\Xhline{0.6pt}
				{\tabincell{c}{Multipath spherical\\wavefont model}} & {\tabincell{c}{NLoS, SM, MAM, MoBM}} & {\tabincell{c}{F1, F3, F5, F6}} & {\tabincell{c}{C1, C3, C4}} & 
				{LCC, LMC, HFMT} & \tabincell{c}{\cite{Cui2022Channel}} & \tabincell{l}{Near-field NLoS channel via the multipath spherical wavefront propagation model\\ and polar-domain channel estimation schemes.} \\
				
				\Xhline{0.6pt}
				{\tabincell{c}{Spatial correlation model}} & {\tabincell{c}{NLoS, SM, MAM, MoBM}} & {\tabincell{c}{F1, F3, F5, F6}} & {\tabincell{c}{C1, C3, C4}} & 
				{LCC, LMC, HFMT} & \cite{Dong2022Near} & \tabincell{l}{Analyze the near-field spatial correlation with an emphasis on the one-ring scatter\\ distribution, including the far-field spatial correlation as a special case.} \\

				\Xhline{0.6pt}
				{\tabincell{c}{Fourier plane-wave\\ series expansion model}} & {\tabincell{c}{NLoS, SM, PCM, MoBM}} & {\tabincell{c}{F1, F3, F5, F6}} & {\tabincell{c}{C1, C3, C4}} & {\tabincell{c}{HA, LCC, LMC,\\ HFMT}} & \tabincell{c}{\cite{Pizzo2022Fourier}} & \tabincell{l}{Present a plane-wave representation of channel response in arbitrary scattering, and\\ provide a low-rank semi-unitarily equivalent approximation of the spatial EM channel.} \\
				\Xhline{0.6pt}
		\end{tabular}}
	\vspace{-1em}
	\end{table*}

\subsection{State-of-the-Art Models}
We next review the state-of-the-art models for near-field H-MIMO propagation channels, comparing them in terms of features, challenges, and evaluation metrics. 
	
\subsubsection{Spherical Wavefont Model \cite{Zhang2022Beam}}
This model describes near-field H-MIMO channels in a mathematically abstracted fashion. It mainly focuses on capturing the spherical wavefront propagation feature while neglecting others, such as the vector wave field and the multiple polarizations. For the LoS component, the amplitude and phase of the channel response between each TX-RX antenna pair are characterized as functions of their distance. 
	
\subsubsection{Parabolic Wavefont Model \cite{Do2023Parabolic}}
This channel model originates from the spherical wavefront propagation model and makes use of a parabolic approximation for each TX-RX element distance. This model facilitates the analysis of LoS channels providing useful insights and properties. However, its applicability requires the antenna apertures to be within a certain size.
	
\subsubsection{Tensor Green's Function Model \cite{Dardari2020Communicating2,Gong2023HMIMO,Gong2023Transmit}}
In the extreme case of H-MIMO, the system will have continuous antenna apertures. To this end, the LoS channel between each TX-RX communication pair can be modeled via the TGF, where the point sources within the propagation environment are within the aperture areas of the TX and RX. This channel model follows the EM wave theory, which is by definition physically consistent.
	
\subsubsection{Multipath Spherical Wavefont Model \cite{Cui2022Channel}}
The near-field H-MIMO channel, which comprises a certain number of propagation paths, is modeled as a sum of the contributions of each propagation path. For each path, the spherical wavefront propagation model is used with a specified distance and angle. 
	
\subsubsection{Spatial Correlation Model \cite{Dong2022Near}}
According to this model, near-field H-MIMO channels are represented by second-order statistics, namely, the spatial correlation matrix. This enables the representation of classes of NLoS channels, with each class including channels  with the same statistics. Each specific realization of the NLoS channel can be generated from the spatial correlation matrix. The near-field spatial correlation matrix is usually derived based on the spherical wavefront propagation channel model.
	
\subsubsection{Fourier Plane Wave Series Expansion Model \cite{Pizzo2022Fourier}} 
This channel model, which is physically consistent, builds on the plane-wave series expansion realized via the Weyl's expansion of the scalar Green's function. It is primarily a far-field NLoS channel model due to the region of the wavenumber domain parameters restricted to the propagation area. When this region is enlarged, near-field wave propagation appears.\footnote{This is the reason why we list this model as a near-field one. In addition, in Table~\ref{tab:ChannelModeling}, we mark this model as HA since it is physically consistent, in contrast to MAMs. However, this model is valid for single polarization, which cannot fully capture the vector wave field.}

\section{Efficient Near-Field H-MIMO Channel Models}
\label{Section_PCM}
In this section, we present the recent near-field H-MIMO LoS channel models of \cite{Gong2023HMIMO,Gong2023Transmit}, which are relevant to the H-MIMO system in Fig.~\ref{fig:ChannelFeatures}(e) and require lower computational and measurement complexities than the available integral form TGF-based
models. 

\subsection{Computationally Efficient EM-Domain Channel Model}
Existing near-field H-MIMO channel models mainly belong to the MAM group, which inherently fails to represent the vector wave field and multiple polarization states. This fact motivated the development of EM-domain PCM. According to the TGF, EM-domain channel modeling is built for theoretically continuous apertures. To model channels for realistic discrete apertures, one may calculate multiple coupled integrals of the TGF over each area of the antenna elements, e.g., the integral form channel model (INTCM) \cite[eq.(5)]{Gong2023HMIMO}, which is however computationally infeasible.
	
To treat the latter feasibility issue, the coordinate-dependent channel model (CDCM) and coordinate-independent channel model (CICM) were proposed in \cite[eqs.(24) and (25)]{Gong2023HMIMO}. These models are integral-free, resulting from expressing multiple coupled integrals via reasonable approximations through Taylor series expansions, and based on an established coordinate representation system. The CDCM can be described as the product of the TX/RX element areas (for rectangle grids), a coordinate-independent TGF, and a coordinate-dependent coefficient. The last term can be further approximated as unity to obtain the CICM. It is worth noting that the underlying assumptions to get these models require the imposed TX current to be uniformly distributed over each TX antenna element and each element radiates EM fields ideally. Compared with the TGF for continuous apertures, CDCM and CICM belong to the same categories (i.e., LoS, DM, PCM, and MoBM), share identical features (F1--F6), face unified challenges (C2--C5), and  have reduced computational complexity. Overall, the evaluation of both CDCM and CICM is HA and LCC.  

\subsection{Measurement Efficient EM-Domain Channel Model}
The TGF-based channel models, including CDCM and CICM, require global position information for the antenna elements and/or all relative distances and directions for all pairs of TX-RX antenna elements. This necessitates large-scale parameter measurements for obtaining the channel coefficients \cite{Gong2023Transmit}, which will inevitably cause practical inconveniences. In addition, the TGF-based models lack of certain flexibility and mathematical tractability compared with MAM, and their scalability to traditional models is limited. 
	
To resolve the latter issue, the partially-separable channel model (PSCM) and its far-field version fully-separable channel model (FSCM) were presented in \cite[eqs.(17) and (25)]{Gong2023Transmit}, which are HA, in comparison with the widely used and mathematically abstracted spherical wavefront channel model, and HFMT. In PSCM, the channel is modeled through a TX-RX parameter separable approach such that the distance between each pair of TX-RX antenna elements can be represented by the following three separable components: surface center distance vector, TX local position vector, and RX local position vector. Notably, only the surface center distance vector needs to be measured in one shot, which reduces significantly the measurement complexity. With this approach, the PSCM is expressed as a Hadamard product of a separable matrix comprising TX/RX array response vectors and an inseparable matrix. On the other hand, FSCM is expressed as functions of the TX/RX array response vectors, where the conventional LoS channel model for M-MIMO is included as a special case. Since PSCM and FSCM are deduced from TGF for discrete apertures, they both follow the same assumptions with those of CDCM and CICM. 

Compared with the TGF-based channel models, PSCM and FSCM belong to the same categories (i.e., LoS, DM, PCM, and MoBM), share identical features (F1--F6), face less challenges than CDCM and CICM (specifically, C3 and C4), and have reduced computational complexity. In terms of evaluation, both PSCM and FSCM are HA, LCC, LMC, and HFMT.

	\begin{table*}[!t]
		\footnotesize
		\renewcommand{\arraystretch}{1.2}
		\caption{\textsc{Computational and Measurement Complexities of the Near-Field H-MIMO LoS Channel Models in \cite{Gong2023HMIMO,Gong2023Transmit}.}}
		\label{Tab:ComputationalMeasurementComplexity}
		\centering
		\resizebox{\linewidth}{!}{
		\begin{threeparttable}
		\begin{tabular}{|c|c|c|c|c|}
			\hline
			Channel model & INTCM \cite[eq.(5)]{Gong2023HMIMO} & \tabincell{c}{CDCM \cite[eq.(24)]{Gong2023HMIMO}\\ \& CICM \cite[eq.(25)]{Gong2023HMIMO}} & \multicolumn{2}{c|}{PSCM \cite[eq.(17)]{Gong2023Transmit} \& FSCM \cite[eq.(25)]{Gong2023Transmit}} \\
			\hline\hline
			Parameters needed & \tabincell{c}{distance vector\\ (each TX-RX grid pair\tnote{\textcolor{red}{*}} )} & \tabincell{c}{distance vector\\ (each TX-RX antenna pair)} & \tabincell{c}{TX/RX local position vectors} & \tabincell{c}{TX-RX aperture\\ center distance vector} \\
			\hline 
			Acquisition method & \multicolumn{2}{c|}{distance/direction measurements} & direct numerical calculations & distance/direction measurements \\
			\hline
			Computational complexity & $\mathcal{O}(3MQNP)$ & $\mathcal{O}(3MN)$ & \multicolumn{2}{c|}{$\mathcal{O}(3MN)$} \\
			\hline
			Measurement complexity & $\ge MQ+NP$ & $\ge M+N$ & N/A & $1$ \\
			\hline
		\end{tabular}
		\begin{tablenotes}
			\footnotesize
			\item[\textcolor{red}{*}] The multiple coupled integrals in INTCM are computed in a discretization manner, where each TX (RX) antenna area is discretized as multiple grid points. 
		\end{tablenotes}
	\end{threeparttable}
	}
	\vspace{-1em}
	\end{table*}

	\begin{figure}[!tbp]
		\centering
		\subfloat[]{\label{fig:WCM_FIG1}\includegraphics[width=0.8\columnwidth]{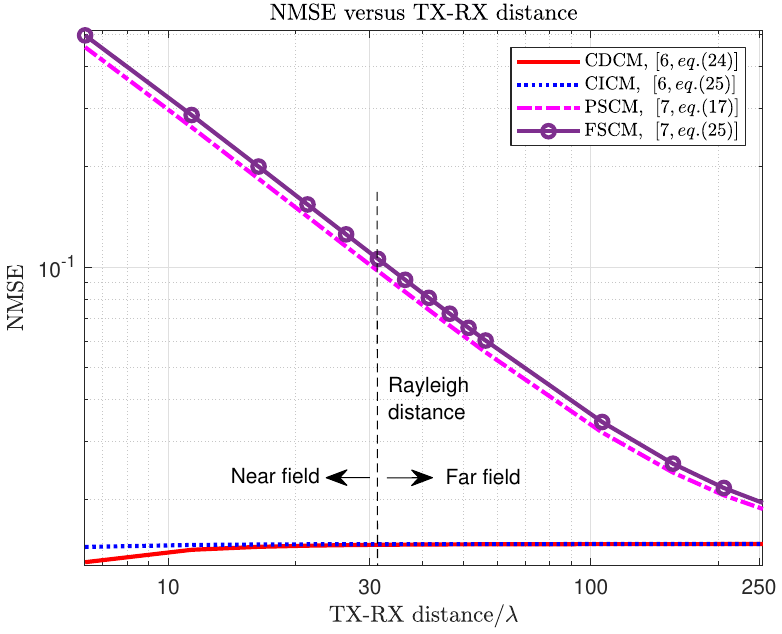}} \\ 
		\subfloat[]{\label{fig:WCM_FIG2}\includegraphics[width=0.8\columnwidth]{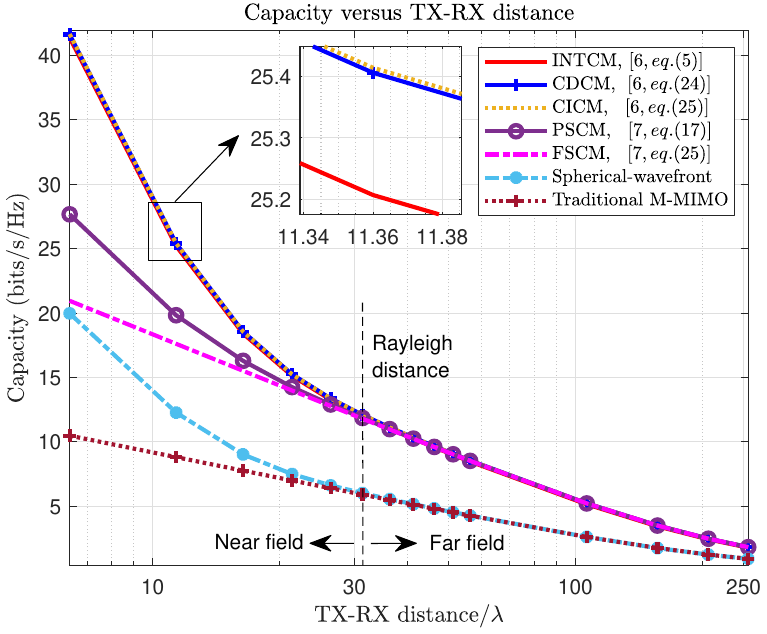}} \\
		\caption{Performance evaluation of the near-field HMIMO LoS channel models in \cite{Gong2023HMIMO,Gong2023Transmit} versus the TX-RX distance: (a) NMSE and (b) achievable spectral efficiency.}
		\label{fig:NMSE_Capacity}
		\vspace{-1em}
	\end{figure}

\subsection{Numerical Results}
We have numerically evaluated the considered CDCM, CICM, and PSCM, as well as the latter's far-field special case FSCM, with respect to the normalized mean squared error (NMSE) and the channel capacity, and compared their computational and measurement complexities. We have also simulated the performance of the models' integral counterpart INTCM. In all simulations, the operating frequency was $2.4$ GHz. The number of antenna elements at TX/RX were chosen as $N = 40 \times 40$ and $M = 16 \times 16$, respectively, and the element spacing was fixed as $0.05 \lambda$\footnote{We have used such a small spacing to retain ultra-high accuracy for the models for demonstration purposes. Such small values can be achieved via advanced technologies, e.g., graphene-based metasurfaces. In addition, the channel models in \cite{Gong2023HMIMO,Gong2023Transmit} consider isotropic radiations without mutual coupling effects; for this, advanced coupling suppression technologies are needed. One may refer to \cite{Chen2018Review} for initial studies of mutual coupling to facilitate coupling-aware channel modeling, as well as various useful coupling reduction technologies.}. The TX/RX apertures were considered to be placed in parallel and their centers were aligned. The tested TX-RX distance varied from $6 \lambda$ to $256 \lambda$. In the capacity evaluation, we have considered average power allocation and fixed the TX signal-to-noise ratio to $20$dB. We also compare the channel models with the spherical wavefront channel model and the far-field LoS channel model used for conventional M-MIMO. 
	
The NMSE performance versus the TX-RX distance is illustrated in Fig.~\ref{fig:WCM_FIG1}. It can be seen that CDCM and CICM attain a quite low NMSE, with the former model outperforming the latter. As the TX-RX- distance increases, the NMSE of CDCM and CICM coincide and get flatten. For PSCM and FSCM, more than $70$\% accuracy is achieved as the distance increases beyond $10 \lambda$. Note that PSCM and FSCM sacrifice accuracy with a significant reduction in measurement complexity. As shown in Table~\ref{Tab:ComputationalMeasurementComplexity}, the computational complexities of CDCM/CICM and PSCM/FSCM are significantly lower than INTCM. The capacity versus the TX-RX distance is depicted in Fig.~\ref{fig:WCM_FIG2}. As observed, both CDCM and CICM results are quite close with those of INTCM. The spherical wavefront model fails to reach the capacity of INTCM, indicating its insufficiency in fully capturing the near-field LoS H-MIMO channel. Interestingly, for the same LCC, LMC and HFMT, PSCM exhibits significantly improved performance than the spherical wavefront model. This can be attributed to its capability to describe the vector wave field. Another noticeable trend is that the near-field models CDCM, CICM, and PSCM degenerate to far-field special cases (i.e., FSCM and the far-field LoS channel model for conventional M-MIMO). One can also observe that FSCM doubles the capacity over that of the far-field LoS channel model for conventional M-MIMO.

\section{Future Research Directions}
	\label{Section_FRD}
Irrespective of the recent significant progress on near-field H-MIMO channel modeling, many open problems still remain unresolved. We next highlight several research directions for future research:
	\begin{itemize}
		\item 
		\textbf{EM-domain near-field NLoS channel modeling.}
The available near-field NLoS channel models are built upon mathematical abstraction indicating that the complex vector wave field and multiple polarizations cannot be fully captured. Additionally, EM-domain near-field channel modeling mainly considers the LoS condition. Therefore, a comprehensive near-field channel model incorporating NLoS conditions and EM principles is required. 
		
		\item 
		\textbf{Channel models with reduced dimensionality.}
Even though the existing near-field H-MIMO channel models can have relatively concise forms, e.g, represented by the array response vectors, the channel dimension is still large due to the extremely large number of antenna elements. Channel models with further reduced dimensionality, amenable to algorithmic designs, are needed.
		
		\item 
		\textbf{Mutual-coupling-aware channel modeling.}
Mutual coupling effects cannot be ignored in a convenient near-field H-MIMO channel model, due to the dense antenna apertures. However, existing models fail to address this fact. It is therefore crucial to understand and accurately model the inherent features of mutual coupling, and then, design efficient mutual-coupling-aware channel models. 
		
		\item 
		\textbf{Spatial-non-stationarity-aware channel modeling.}
Spatial non-stationarity effects will be also present due to the large H-MIMO antenna apertures. However, existing near-field H-MIMO channel models do not consider them. To this end, scatterer-aware designs and spatial-non-stationarity-aware channel models are important research directions for the topic. 
		
		\item 
		\textbf{Channel sounding and validations.}
Realistic channel measurements for near-field H-MIMO constitutes a task of great significance. Those measurements will enable the validation of theoretical models for various near-field H-MIMO parameters, for example, propagation delays, path losses, angular spreads, and polarization.
	\end{itemize}

	\section{Conclusion}
	\label{Section_CON}
In this article, we focused on near-field channel modeling for emerging H-MIMO systems. We first presented the main categories of existing channel models and then described several existing near-field H-MIMO channel models in terms of their modeled channel features, challenges, and evaluation criteria. To tackle the high computational and measurement complexity of TGF-based LoS channel models, we presented two groups of highly accurate near-field H-MIMO channel models, the one being computationally efficient and the other measurement efficient. The article was concluded with a list of future research directions, which pave the way for more precise and efficient near-field channel models to empower future H-MIMO wireless systems.

    \section*{Acknowledgment}
    The work was supported in part by the Ministry of Education, Singapore, under its MOE Tier 2 (Award number MOE-T2EP50220-0019), and in part by the Science and Engineering Research Council of A*STAR (Agency for Science, Technology and Research) Singapore, under Grant No. M22L1b0110. 
    The work of Chongwen Huang was supported in part by the China National Key R\&D Program under Grant 2021YFA1000500 and 2023YFB2904800, in part by National Natural Science Foundation of China under Grant 62331023, 62101492, 62394292 and U20A20158, in part by Zhejiang Provincial Natural Science Foundation of China under Grant LR22F010002, in part by Zhejiang Provincial Science and Technology Plan Project under Grant 2024C01033, and in part by Zhejiang University Global Partnership Fund. 
    The work of George C. Alexandropoulos was supported by the Smart Networks and Services Joint Undertaking (SNS JU) project TERRAMETA under the European Union's Horizon Europe research and innovation programme under Grant Agreement no. 101097101, including top-up funding by UK Research and Innovation (UKRI) under the UK government's Horizon Europe funding guarantee.


	\bibliographystyle{IEEEtran}
	\bibliography{IEEEabrv,references} 

    
    \begin{IEEEbiographynophoto}{Tierui Gong}
        (S'18-M'20) received the Ph.D. degree from University of Chinese Academy of Sciences (UCAS), Beijing, China, in 2020. 
        He is currently a Research Fellow with the School of Electrical and Electronic Engineering, Nanyang Technological University (NTU), Singapore. His research interests include holographic MIMO communications, electromagnetic signal and information theory, massive MIMO communications, and signal processing for communications. 
    \end{IEEEbiographynophoto}

    \begin{IEEEbiographynophoto}{Li Wei}
        received the B.Sc. degree in communication engineering from Southwest Jiaotong University, China, in 2015, the M.Sc. degree in electronic and communication engineering from Xidian University, China, in 2019, and the Ph.D. degree from Singapore University of Technology and Design, Singapore, in 2023. She is currently a Research Fellow with the School of Electrical and Electronic Engineering, Nanyang Technological University, Singapore. Her research interests include holographic MIMO communications, reconfigurable intelligent surfaces, etc.
    \end{IEEEbiographynophoto}

    \begin{IEEEbiographynophoto}{Chongwen Huang}
        obtained his B.Sc. degree in 2010 from Nankai University, the M.Sc degree from the University of Electronic Science and Technology of China in 2013, and the PhD degree from Singapore University of Technology and Design (SUTD) in 2019. Since Sep. 2020, he joined into Zhejiang University as a tenure-track young professor. His main research interests are focused on Holographic MIMO Surface/Reconfigurable Intelligent Surface, B5G/6G Wireless Communications, mmWave/THz Communications, Deep Learning technologies for Wireless communications, etc.
    \end{IEEEbiographynophoto}

    \begin{IEEEbiographynophoto}{George C. Alexandropoulos}
        (S'07-M'10-SM'15) is an Associate Professor in the Department of Informatics and Telecommunications, National and Kapodistrian University of Athens Greece. His research interests span the general areas of algorithmic design and performance analysis for wireless networks with emphasis on multi-antenna transceiver hardware architectures, active and passive metasurfaces, full-duplex radios, and millimeter-wave/THz communications, as well as distributed machine learning algorithms. He has participated and/or technically managed numerous EU, international, and Greek research, innovation, and development projects, including H2020 RISE-6G, ESA PRISM, SNS JU TERRAMETA, and SNS JU 6G-DISAC.
    \end{IEEEbiographynophoto}

    \begin{IEEEbiographynophoto}{Mérouane Debbah}
        (S’01-M’04-SM’08-F’15) received the M.Sc. and Ph.D. degrees from the Ecole Normale Supérieure Paris-Saclay, France. Since 2023, he is a  Professor at  Khalifa University of Science and Technology in Abu Dhabi and founding director of the 6G center. He has managed 8 EU projects and more than 24 national and international projects. His research interests lie in fundamental mathematics, algorithms, statistics, information, and communication sciences research. He holds more than 50 patents. He is an IEEE Fellow, a WWRF Fellow, a Eurasip Fellow, an AAIA Fellow, an Institut Louis Bachelier Fellow and a Membre émérite SEE. 
    \end{IEEEbiographynophoto}

    \begin{IEEEbiographynophoto}{Chau Yuen}
	(S'02-M'06-SM'12-F'21) received the B.Eng. and Ph.D. degrees from Nanyang Technological University, Singapore, in 2000 and 2004, respectively. Since 2023, he has been with the School of Electrical and Electronic Engineering, Nanyang Technological University. 
	He is a Distinguished Lecturer of IEEE Vehicular Technology Society, Top 2\% Scientists by Stanford University, and also a Highly Cited Researcher by Clarivate Web of Science. He has 3 US patents and published over 500 research papers at international journals or conferences.
    \end{IEEEbiographynophoto}

\end{document}